# Long Term Operation and Performance of Cryogenic Sapphire Oscillators

Tobar M.E.[1], Ivanov E.N.[1], Locke C.R.[1], Stanwix P.L.[1], Hartnett J.G.[1], Luiten A.N.[1], Warrington R.B.[2], Fisk P. T. H.[2], Lawn M.A.[2], Wouters M.J.[2], Bize S.[3], Santarelli G.[3], Wolf. P.[3,4], Clairon A.[3], Guillemot P.[5]

[1]Frequency Standards and Metrology group, School of Physics University of Western Australia, Crawley, WA 6009 Australia

[2]National Measurement Institute Bradfield Road, West Lindfield, NSW 2070 Australia

[3]LNE-SYRTE, 61, Avenue de l'Observatoire, 75014 Paris, France.

[4] Bureau International des Poids et Mesures, Pavillon de Breteuil, 92312 Sevres Cedex, France.

[5]CNES, French Space Agency, Toulouse, France.

*Abstract*—**Cryogenic Sapphire Oscillators (CSO) developed at UWA have now been in operation around the world continuously for many years. Such oscillators, due to their excellent spectral purity are essential for interrogating atomic frequency standards at the limit of quantum projection noise; otherwise aliasing effects will dominate the frequency stability due to the periodic sampling between successive interrogations of the atomic transition. For this reason, UWA oscillators are now operational at NMI (Sydney), LNE-SYRTE (Paris), the French Space Agency (CNES, Toulouse) and at UWA (Perth). Other applications, which have attracted attention in recent years, include tests on fundamental principles of physics, such as tests of Lorentz invariance. This paper reports on the long-term operation and performance of such oscillators. We compare the long-term drift of some different CSO's. The drift rates turn out to be linear over many years and in the same direction. However, the magnitude seems to vary by more than one order of magnitude between the oscillators, ranging from $10^{14}$ per day to a few parts in $10^{13}$ per day.**

I. INTRODUCTION

The University of Western Australia's (UWA) Cryogenic Sapphire Oscillator (CSO) projects have a rich history of development, which has involved many postdoctoral researchers and Ph.D. students since 1982. Inspired by the early work of Braginsky [1, 2], the first work at UWA on sapphire Whispering Gallery (WG) mode resonators began in 1982 [3]. The main drive for developing such devices at UWA was the proposition to develop low noise oscillators for use in microwave displacement transducers for Gravitational Wave detection and measurement of the Standard Quantum Limit (~ 1 kHz from carrier) [4-9]. Following this, work began on ultra-frequency-stable oscillators for periods greater than one second based on the CSO, which is the topic of this work. So far the UWA developed CSO has undergone three major redesigns since 1989, and are referred to as first-, second- and third-generation designs in historical order.

The first successful development of the CSO was achieved in 1989 (first-generation), which exhibited frequency instability better than one part in $10^{14}$ ($9 \times 10^{-15}$ at 10 to 300 seconds) [12, 13]. At the heart of the CSO was a sapphire cylinder of 3 cm diameter placed in a superconducting niobium cavity. At that point in time the structure of the WG modes within the resonators were unknown. The mode structures were first determined during 1990 by solving Maxwell's equations in uniaxial anisotropic dielectric resonators, and applying to an open structure [11]. Later, the problem including the enclosing cavity was solved [12].

One of the first-generation sapphire clocks was transported to the National Measurement Laboratory (now the National Measurement Institute, NMI) during 1993. The oscillator was used as the flywheel oscillator for a microwave atomic clock based on trapped, laser-cooled ytterbium ions. This clock was the first trapped ion clock to combine the extreme accuracy possible achieved by cooling the ions to sub-Kelvin temperatures, thus reducing relativistic frequency shifts due to ion motion to less than 1 part in

$10^{15}$, with the improved signal-to-noise ratio possible by using a relatively large number ($10^4$) of ions. This combination of accuracy and frequency stability was unique in a trapped ion atomic clock, and enabled the NMI clock to have performance comparable to that of a cesium fountain clock [13]. The oscillators have also recently been implemented in a new rotating experiment to test Lorentz Invariance [14-17].

To increase the Q-factor and reduce the cavity wall effect on the resonance frequency, larger pieces of sapphire (5cm diameter rather than 3cm) were implemented in the second-generation CSO's, which allowed the frequency instabilities to be reduced by a factor of 3, to between 3 and $4 \times 10^{-15}$ at 0.2 to 100 seconds [18]. During 1998 a second-generation sapphire oscillator was transported from UWA to SYTRE (Paris) for 1 month. This enabled the best performance of the SYRTE fountain at the Quantum Projection Noise (QPN) limit [19]. Later, further optimization of the electronics and sapphire resonator enabled a further order of magnitude improvement in performance of $2.4 \times 10^{-16}$ at 32 seconds [20].

During 2000 an improved second-generation CSO was transported again to SYRTE at the Paris Observatory. The oscillator allowed some new tests on Lorentz Invariance, up to a factor of 70 times more superior than previously [21-23]. The oscillator now supplies the stable frequency used to synthesize the frequencies that excite atomic transitions of the atomic clocks at SYRTE [24,25].

During 2002 we set out to redesign the sapphire oscillators so they were easier to transport from UWA to another site (third-generation). This was prompted by the requirement of CNES to have such an oscillator to enable ground tests on the PHARAO atomic space clock, for the Atomic Clock Ensemble in Space (ACES) mission. The oscillator was transported to CNES in Toulouse during 2003. The short-term frequency stability was measured and limited by the hydrogen maser used in the comparison ($<4 \times 10^{-15}$ at $10^2$-$10^3$ seconds) [26]; the long-term results are reported in this paper.

The short-term performance of various configurations of the CSO's has been well characterized previously and currently represents the state-of-the-art in frequency stability. In this work we concentrate on comparing the long-term performance of the UWA designed CSO's for periods of multiple years, starting from the latest third-generation oscillators wherein we also introduce the new design changes implemented.

Finally we compare results of the UWA designed CSO's with that obtained by other groups.

II. Third-Generation CSO At CNES (French Space Agency)

A. *New cryogenic insert*

A third-generation CSO now resides at CNES in Toulouse [27]. The main design change implemented was that of the cryogenic insert that houses the sapphire whispering gallery mode cavity, which we describe here. Fig. 1 shows a schematic of the inner can and outer vacuum chambers that supports the cavity. The niobium cavity encasing the sapphire resonator is immersed in a helium bath (~4K) to reduce thermal noise and increase the Q-factor of the resonance. The helium bath is shielded from the environment by a heavily insulated dewar approximately 1.8m high and 0.5m wide. The sapphire itself is attached to the lower end of an insulating stainless steel "insert" which is placed into this dewar. The insert consists of a series of baffles to reduce convection and maintain temperature within the dewar. Stainless steel microwave transmission lines are used to complete the room temperature to liquid helium section of the oscillator circuit due to their low thermal conductivity. To prevent stress on the microwave fitting associated with thermal cycling, loops are placed within the microwave cables.

While it is important to maintain the resonator at helium temperatures, it is also important to thermally isolate it from the surrounding helium bath to reduce the severity of any temperature fluctuations. As such, the niobium cavity is contained within an "inner can" turbo pumped to ~$10^{-6}$ Torr at room temperature, and then further cryo-pumped upon cooling. Heat is extracted from this inner can via the copper-mounting shaft. The inner can is contained within a second "outer can" which is pumped to ~$10^{-3}$ Torr to reduce thermal conduction into the dewar and also to provide a vacuum environment for the microwave components, which sit inside the can. Heat is drawn from the outer can via three copper heat sinks, which sit in the helium bath.

A heater wire is situated on the copper post inside the outer can to maintain the sapphire at its turning point in the frequency/temperature dependence (around 6 K). Calibrated Carbon-Glass and Germanium RTDs situated on the copper post in the outer can were used to measure the temperature. The turning point is due to $Ti^{+3}$ and $Mo^{+3}$ ions and has a curvature of 3 Hz $K^{-2}$. Thus, to achieve fractional frequency stability of order $10^{-16}$ at 1 mK offset from the turning point, a temperature stability of 0.3 mK would be necessary.

In previous generations, there were problems experienced with the way we created vacuum seals. The seals were achieved by crushing indium wire within a tongue and groove. Leaks were common and difficult to find, and the whole process needed to be repeated when thermally cycling the experiment. A new Mylar seal system was developed, which may be cycled repeatedly. The seals are easily cleaned and recycled when compared to our previous indium system. All microwave and wiring feed-throughs have been made as modular units for ease of assembly and to allow more flexibility with placement. The new seals were found to be most effective and no leaks have been detected in any of the seals implemented. The inner can itself was redesigned as a smaller, lighter unit. A rounded bottom maintains structural rigidity while allowing for thinner walls. It also reduces the amount of liquid helium required to initially cool the can.

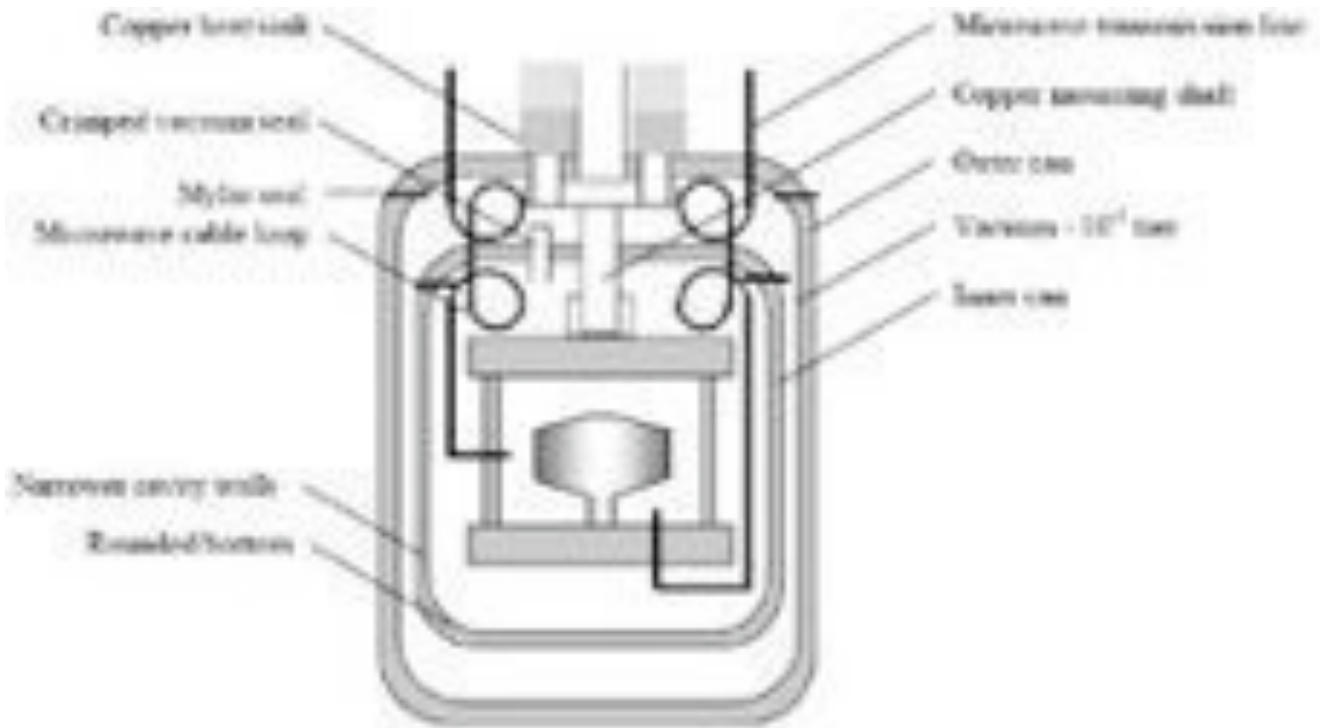

Figure 1. The inner and outer can construction.

B. *Long-term performance of the CNES CSO*

The CSO was transported to CNES to act as the local oscillator for the engineering model of the PHARAO space clock [22]. Also, during the mission it is planned to operate as part of the ground station ensemble of stable oscillators and clocks. To measure the frequency stability the CSO, it is compared to the CNES hydrogen maser. Below 300 seconds the measurement is limited by the maser, above 300 seconds it is limited by the CSO. For details on how the measurement is achieved see reference [22]. The fractional frequency shift of the CSO has been recorded from 2003 to 2005 and is plotted in Fig 2. During cryogenic cycling there was an abrupt frequency jump of about 0.5 Hz, however the drift otherwise remain at a constant level of $-2.4\times10^{-13}$/day. The frequency jump is small compared to the bandwidth of the sapphire resonator, which is of the order of 10 Hz and most likely arises due to irreversible changes to the resonator through cycling to room temperature. For example, this could occur due to a shift in coupling or cavity dimension. Whatever the cause it does not impinge on the operation of the oscillator as the frequency shift is quite small.

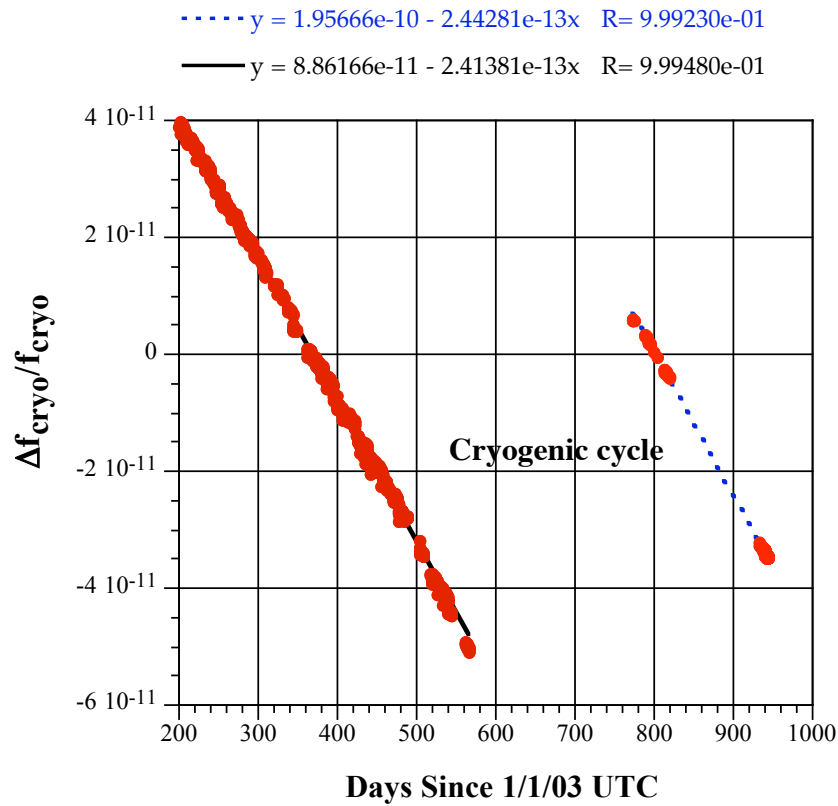

Figure 2. CSO frequency versus time in days since 1/1/03 UTC. The drift rate is linear at $-2.4 \times 10^{-13}$/day. Frequency data is shown by the clusters of solid dots, and curve fits are shown by the thin solid and dashed lines. The drift remains constant and linear even with the oscillator electronic circuit is off, which is shown by the periods of no data. After cryogenic cycling, there is a frequency jump, however the linear drift remains constant.

### III. CSO AT THE PARIS OBSERVATORY

The second-generation CSO at the Paris Observatory is well described in references [19-24]. It serves as a master oscillator of a frequency synthesis chain providing a spectrally clean microwave signal for probing atomic transitions at the QPN limit. To correct for the relatively large long-term drift of the CSO, the oscillator is weakly phase locked to a hydrogen maser (Fig. 3). It thus has the stability of the CSO at short times ($<10^{-15}$ up to 800 seconds) and the hydrogen maser over the long-term. This data has been used previously to test the constancy of the speed of light and the photon sector of the Standard Model of Physics, which rely on the Earth's rotation and orbit [21-23]. In this work we use the same data to analyze the long-term performance of the CSO.

The final synthesized beat frequency between the hydrogen maser and the CSO has drifted between 65 to 63 Hz since 2002. The fractional beat frequency is plotted against time in Fig.4. The striking feature is the constant linear drift of $-1.5 \times 10^{-13}$/day over a period of more than three years. As for the CNES CSO there is a frequency shift during cycling. However, the drift remains constant at a similar level.

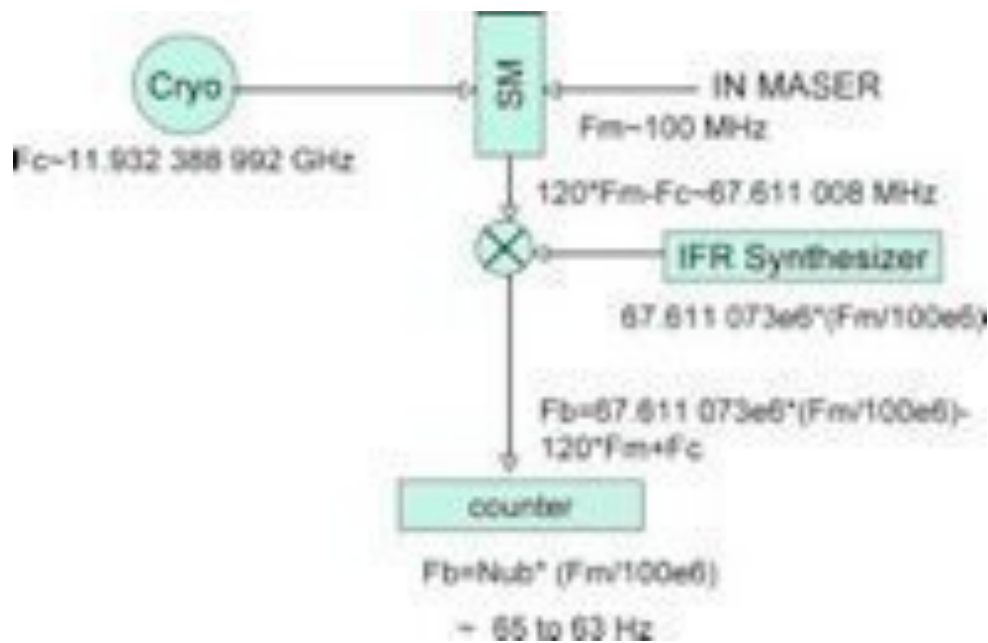

Figure 3. Schematic of the synthesized comparison between the the CSO (Cryo) with respect to a Hydrogen Maser [24]. The resulting beat frequency is between 65 to 63 Hz. Here SM is a sampling mixer.

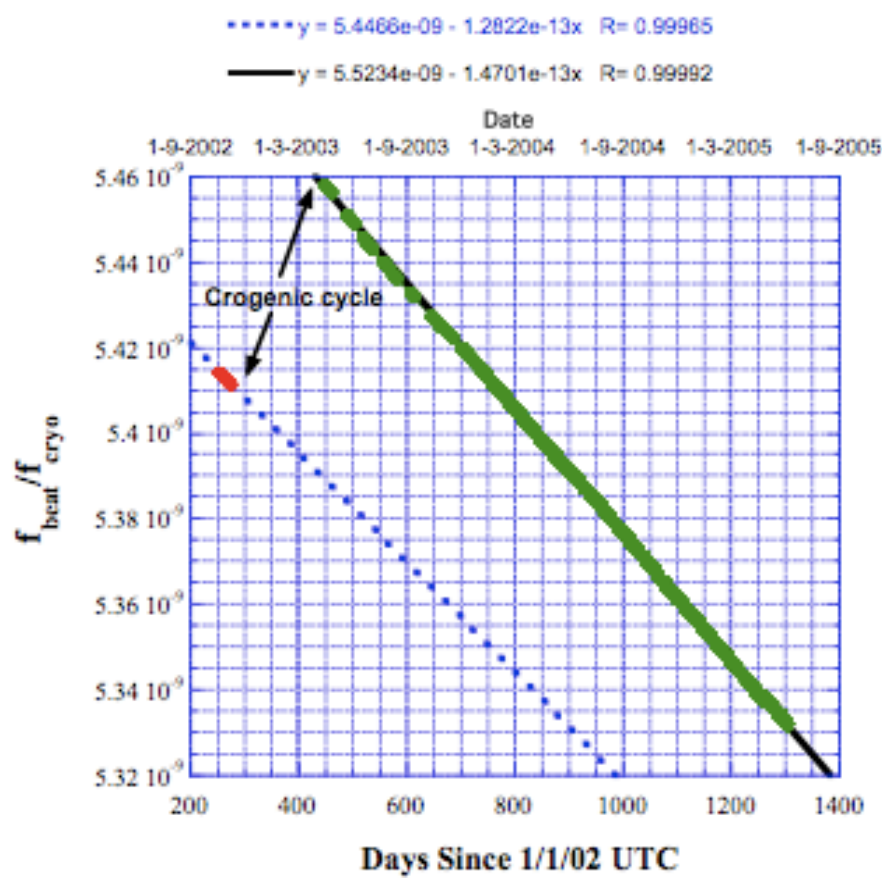

Figure 4. Fractional beat frequency with respect to the SYRTE CSO versus time in days. Between 2002 and 2003 the oscillator was warmed from 4 K to room temperature for maintainence, resulting in a frequency shift. However, the drift rate remained similar and of the order of $-10^{-13}$/day. In the periods where cryogenic cycling did not take place the drift remained linear and constant even between periods when the oscillator was turned off. The thick solid lines are actual data, while the dashed line is the extended curve fit for the data taken in 2002, and the thin solid line is the curve fit for data take after 2003.

## IV.  CSO AT THE NATIONAL MEASUREMENT INSTITUTE

A first-generation CSO was transported to the National Measurement Institute (NMI) in Sydney 1993. The oscillator has been in continuous operation for the last ten years. To our knowledge this is the longest continuous operation without cycling of such a system. The frequency is measured with respect to a hydrogen maser as shown in Fig. 5.

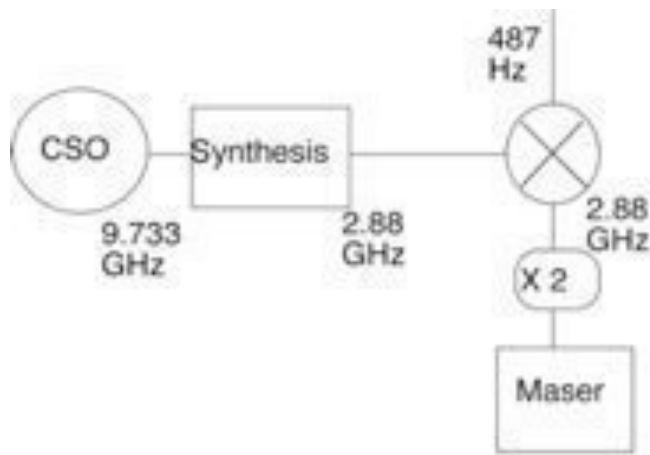

Figure 5. Schematic of the comparison of the CSO with a hydrogen maser at NMI. The beat frequency at 487 Hz is monitored and plotted over ten years in Fig. 6.

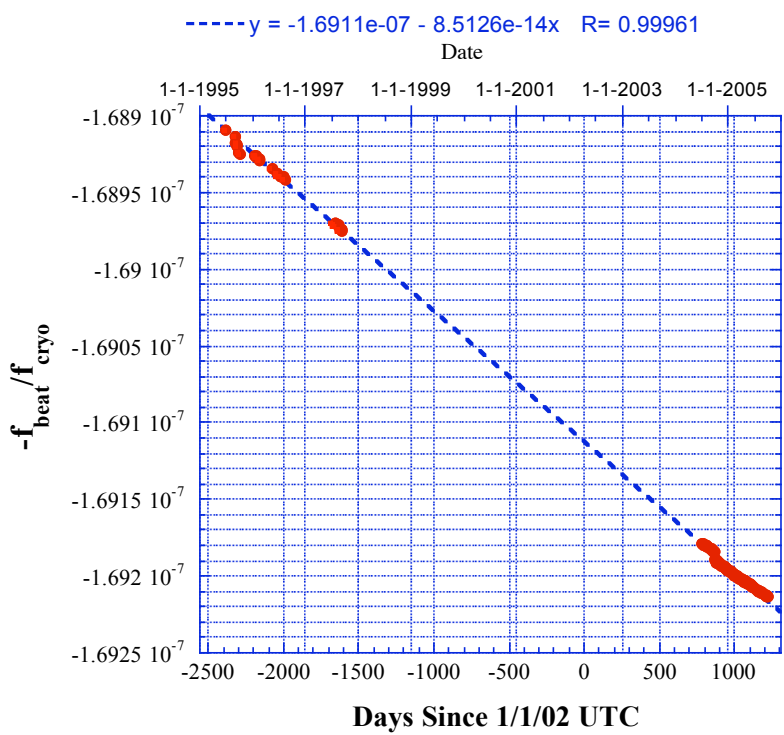

Figure 6. Fractional beat frequency between the CSO and hydrogen maser as a function of time (days since 1/1/02 UTC). The CSO was not cycled cryogenically at all during this period. The drift rate is linear and corresponds to the CSO frequency changing by a fraction of $-8.5 \times 10^{-14}$/day. The precision of the curve fit is not improved with an exponential fit, which shows that the drift is linear for at least ten years. Also, the drift rate remains constant and linear irrespective of changes in the oscillators electronic configuration and periods of down time, which suggests the drift is not due to the electronic circuit.

The beat frequency with respect to the CSO frequency is plotted in Fig. 6. Because the 2.88 GHz synthesized from the CSO is less than the 2.88 GHz synthesized from the maser, a decrease in CSO frequency is actually recorded as an increase in the beat frequency. A constant linear drift of $-8.5\times10^{-14}$/day is observed over that period.

To determine whether the frequency-temperature turning point characteristics change with time it was re-measured during May 2004 to compare with an earlier measurement during January 1994. This was in case any of the frequency drift was actually due to a drift in the temperature sensor used for the temperature stabilization. We measured only a difference of 20mK in the turnover, which would contribute only to a couple of mHz frequency offset, which is negligible to the net change of 0.8 Hz since 1995.

## V.  COMPARISON OF LONG-TERM PERFORMANCE

Comparing the results discussed above with that of other research groups (discussed in this section) we postulate that the limit of the long-term performance of the UWA oscillators is governed by stress release at the resonator supports, and present evidence based on our work and others. The likely mechanism of frequency change is due to stressed induced changes of permittivity (which have been measured previously to be a larger effect than dimensional changes) [28]. In the original 1$^{st}$ generation UWA design the resonators were held tightly from both ends as shown in figure 7. Nevertheless the oscillator has excellent long-term performance of $-8.5\times10^{-14}$/day. Initial the second-generation resonators (with sloped sides as shown in figure 8) were designed with the same support structure. However, it was found by Chang et. al. [29] that the long-term drift was degraded and had a magnitude of greater than parts in $10^{12}$/day. The degraded performance is likely due to the sloped sides providing a path for stress to be transferred from the support to the outer part of the resonator where the Whispering Gallery modes exist. To reduce the amount of stress, the support was changed to a structure that grips one of the support spindles and leaves the other free. This reduced the frequency drift by one order of magnitude to the order of parts in $-10^{-13}$/day, and now all oscillators with these resonators observe about this value of drift (CSO's at Paris and Toulouse). Because this result is larger value than the UWA oscillator at NMI (about a factor of 2 or more), the results suggests that the resonators with sloped sides may be still exhibiting frequency drift due to stress at the support translating to sapphire crystal. It is also likely that the frequency drift of the NMI oscillator is due to stress release as the resonator is over constrained by being clamped between both ends. By gripping the NMI resonator at one end of the resonator instead of two, the drift of the NMI oscillator could probably be further reduced.

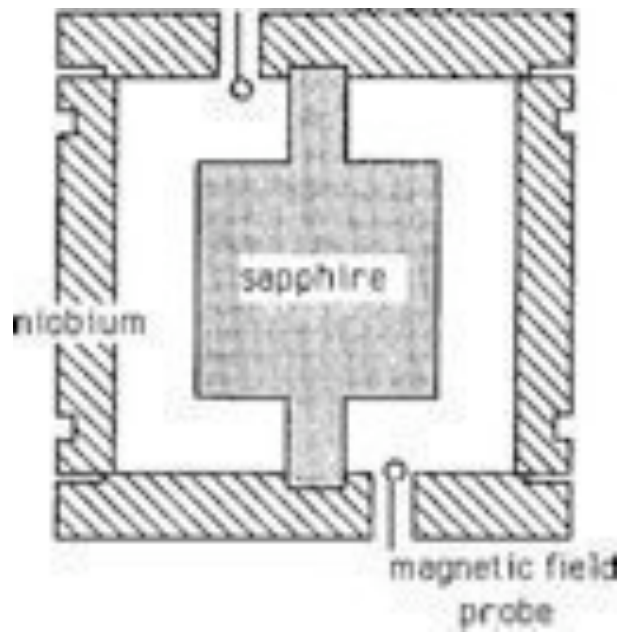

Figure 7. First generation resonator of 3 cm diameter [12,13], which is clamped and stressed at both ends of the sapphire spindle.

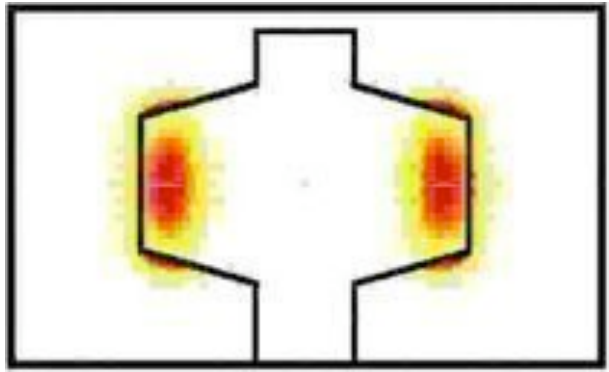

Figure 8. Schematic of the sapphire resonators for both the SYRTE and CNES oscillators, which have sloped sides, and now are supported by one end [29], reducing the stress induced frequency shift in both oscillators to the order of $-10^{-13}$/day.

The group at the Laboratoire de Physique et Métrologie des Oscillateurs (LPMO) have developed a CSO made from a 5 cm sapphire cylinder which is supported at one end in an open configuration (no outer cavity) [30, 31]. The drift is a further factor of 4 smaller than the UWA oscillator at NMI. One would expect a reduced stress induced drift of the LPMO oscillator due to the larger radius of the sapphire resonator and the fact that it is only gripped at one end. However, whether or not their results are still limited by stress release is an open question. The best long-term results in the literature have been obtained by the group at the Jet Propulsion Laboratory (JPL) [32], of $-10^{-14}$/day. They have taken considerable care in reducing the stress induced at the support structure [33]. The results presented in this paper support the design methodology of JPL, if one is to maintain a small long-term drift rate of such oscillators. Other factors that might influence the drift rate that have not been studied include mechanical sagging, slow vacuum changes, cavity effects or effects of nearby spurious modes. It is possible that long-term drift in the JPL and LPMO oscillator may be limited by such effects. However, it does seem that the drift in the UWA oscillators are still limited by stress release at the supports, due to the degraded performance with respect to the other oscillators compared in this work. Nevertheless, an interesting result is that all these oscillators maintain a linear drift for a long period. If one fits the data with an exponential the precision of the curve fits are worse or the same as the linear curve fits.

ACKNOWLEDGMENT

This work was funded by the Australian Research Council.

REFERENCES

[1] V. Braginsky, V. Panov, "Superconducting resonators on sapphire," IEEE Trans. Magn., vol. 15, no. 1, pp. 30-32, 1979.

[2] V. Braginsky, V. Panov, S. Vasiliev, "The properties of superconducting resonators on sapphire," IEEE Trans. Magn., vol. 17, no. 1, pp. 30-32, 1981.

[3] D.G. Blair and I.N. Evans, "High-Q microwave properties of a sapphire ring resonator," J. Phys. D: Appl. Phys., vol. 15, no. 9, 1651-165, 1982.

[4] M.E. Tobar, E.N. Ivanov, R.A. Woode, J.H. Searls, A.G. Mann, "Low noise 9 GHz sapphire resonator-oscillator with thermoelectric temperature stabilisation at 300 Kelvin," IEEE MWGL, vol 5, no. 4, pp. 108 - 110, 1995.

[5] M.E. Tobar and D.G. Blair, "Phase noise analysis of the sapphire loaded superconducting niobiumcavity oscillator," IEEE Trans. Micrw. Theory Tech., pp. 344-347, 1994.

[6] R.A. Woode, M.E. Tobar, E.N. Ivanov, D.G. Blair, "An ultralow noise microwave oscillator based on a high-Q liquidnitrogen cooled sapphire resonator," IEEE Trans. Ultrason. Ferroelec. Freq. Contr., vol. 43, no. 5, pp. 936 - 941, 1996.

[7] E.N. Ivanov, M.E. Tobar, R.A. Woode, "Microwave interferometry: application to precision measurements andnoise reduction techniques," IEEE Trans. Ultrason. Ferroelec. Freq. Contr., vol. 45, no. 6, pp. 1526-1536, 1998.

[8] ME Tobar, EN Ivanov, D Oi, BD Cuthbertson, DG Blair, "Sapphire test-masses for measuring the standard quantum limit and achieving quantum non-demolition," Applied Physics B, vol. 64, pp. 153-166, 1997.

[9] DG Blair, EN Ivanov, ME Tobar, PJ Turner F Van Kann and IS Heng, "High sensitivity gravitational wave antenna with parametric transducer readout," Phys. Rev. Let., vol. 74, no. 11, pp. 1908-1911, 1995.

[10] A.J. Giles, A.G. Mann, S.K. Jones, D.G. Blair, M.J. Buckingham. "A very high stability sapphire loaded superconducting cavity oscillator," Physica B: Condensed Matter, vol. 165, pp. 145-146, 1990.

[11] M.E. Tobar and A.G. Mann, "Resonant frequencies of higher order modes in cylindricalanisotropic dielectric resonators," IEEE Trans. Microw. Theory Techn., vol. 39, no. 12, pp. 2077-2082, 1991.

[12] E.N. Ivanov, D.G. Blair, V.I. Kalinichev, "Approximate approach to the design of shielded dielectric disk resonators with whispering gallery modes." IEEE Trans. MTT, vol. 41, no. 4, pp. 632 - 638,1993.

[13] P.T.H. Fisk, M.J. Sellars, M.A. Lawn, C. Coles, A.G. Mann, D.G Blair, 'Very high Q microwave spectroscopy on trapped $171Yb^+$ ions: application as a frequency standard," IEEE Trans. on Instrum. Meas., vol. 44, no. 2, pp. 113-116, 1995.

[14] PL Stanwix, ME Tobar, P Wolf, M Susli, CR Locke, EN Ivanov, J Winterflood, F van Kann, "Test of Lorentz invariance in electrodynamics using rotating cryogenic sapphire microwave oscillators," Phys. Rev. Lett., vol. 95, 040404, 2005.

[15] P.L. Stanwix, M.E. Tobar, J. Winterflood, E.N. Ivanov, M. Susli, J.G. Hartnett, F. van Kann, P. Wolf "Rotating Michelson-Morley experiment based on a dual-cavity cryogenic sapphire oscillator,"

in Proc 2004 IEEE International Ultrasonics, Ferroelectrics, and Frequency Control 50th Anniversary Joint Conference, pp. 757-761.

[16] M.E. Tobar, P. Wolf, P.L. Stanwix, A. Fowler, S. Bize, A. Clairon, J.G. Hartnett, E.N. Ivanov, F. van Kann, G. Santarelli, M. Susli, J. Winterflood. "New tests of Lorentz Invariance in the Photon Sector using Precision Oscillators and Interferometers," in *CPT AND LORENTZ SYMMETRY Proceedings of the Third Meeting Bloomington,* USA 4 - 7 August 2004, Editor: V Alan Kostelecký, World Scientific pp. 20-28, 2005.

[17] M.E. Tobar, P.L Stanwix, M. Susli, P. Wolf, C.R. Locke, E.N. Ivanov, "Rotating Resonator-Oscillator Experiments to Test Lorentz Invariance in Electrodynamics," Editors: Jurgen Ehlers and Claus Lammerzahl, to be published in Springer Verlag, Lecture notes in Physics, ArXiv: hep-ph/0506200, 2005.

[18] A.N. Luiten A.G. Mann, M.E. Costa, D.G. Blair, "Power Stabilized Cryogenic Sapphire Oscillator," IEEE Trans. Instr. Meas., vol. 44, no. 2, pp. 132-135, 1995.

[19] G. Santarelli, Ph. Laurent, P. Lemonde, A. Clairon, A.G. Mann, S. Chang, A. N. Luiten, C. Salomon, "Quantum projection noise in an atomic fountain: A high stability cesium frequency standard," Phys. Rev. Lett., vol. 82, no. 23, pp. 4619–4622, 1999.

[20] S. Chang, A.G. Mann, A.N. Luiten, "Improved cryogenic sapphire oscillator with exceptionally high frequency stability," Electron. Lett., vol. 36, no. 5, pp. 480 - 481, 2000.

[21] P Wolf, S Bize, A Clairon, AN Luiten, G Santarelli, ME Tobar, "Tests of Lorentz Invariance using a microwave resonator," Phys. Rev. Lett., vol. 90, no. 6, 060402, 2003.

[22] P Wolf, ME Tobar, S Bize, A Clairon, AN Luiten, G Santarelli, "Whispering gallery resonators and tests of Lorentz invariance," GRG, vol. 36, no. 10, pp. 2351-2373, 2004.

[23] P Wolf, S Bize, A Clairon, G Santarelli, ME Tobar, AN Luiten, "Improved test of Lorentz invariance in electrodynamics," Phys. Rev. D, vol. 70, no. 5, 051902, 2004.

[24] D. Chambon, et. al., "Design and realization of a flywheel oscillator for advanced time and frequency metrology," Rev. Sci. Instrum., vol. 76, 094704, 2005.

[25] C. Vian, et.al. "BNM-SYRTE Fountains: Recent Results," IEEE Trans. Instrum. Meas., vol. 54, no. 2, pp. 833-836, 2005.

[26] P. Guillemont, et. al., "The PHARAO time and frequency performance verification system," Proc. of the IEEE Int. Freq. Contr. Symp., pp. 785-789, 2004.

[27] C.R. Locke, S. Munro, M.E. Tobar, E.N. Ivanov, G. Santarelli, Proc. of the Joint Meeting EFTF/IFCS, pp. 350-354, 2003.

[28] CR Locke, ME Tobar, "Measurement on strain induced coefficient of permittivity of sapphire using whispering gallery modes excited in high-Q acoustic sapphire oscillators," Meas. Sci. Technol., vol. 15, pp 2145-2149, 2004.

[29] S. Chang, A.G. Mann, "Mechanical stress caused frequency drift in cryogenic sapphire oscillators," Proc. of the IEEE Int. Freq. Contr. Symp., pp. 710-714, 2001.

[30] P.Y. Bourgeois, F. Lardet-Vieudrin, Y. Kersalé, N. Bazin, M. Chaubet and V. Giordano, "Ultra low drift microwave cryogenic oscillator," Electron. Lett., vol. 40, no. 10, pp 605-606, 2004.

[31] P.Y. Bourgeois, "Reference secondaire de frequence a resonateur saphir cryogenique," Ph.D. Thesis, Univerité de Franche-Comté, 2004.

[32] R.T. Wang , G.J. Dick, and W.A. Diener, " Progress on 10 kelvin cryo-cooled sapphire oscillator", Proceedings of the 2004 IEEE International Frequency Control Symposium, pp. 752-756, 2004.

[33] G.J. Dick and R.T. Wang, "Cryo-cooled sapphire oscillator operating above 35K," Proc. 2000 International IEEE Frequency Control Symposium, 480-484, 2000.